\documentclass[11pt]{article}
\usepackage{moriond,epsfig}
\usepackage{axodraw}

\def\be{\begin{equation}}
\def\ee{\end{equation}}
\def\bea{\begin{eqnarray}}
\def\eea{\end{eqnarray}}

\begin{document}
\rightline{LPT--Orsay 08-49}
\vspace*{4cm}
\title{{\textbf{ $\ell_1 \rightarrow \ell_2 \gamma$}} in type III seesaw}

\author{ F. BONNET \footnote{florian.bonnet@th.u-psud.fr}}

\address{Laboratoire de Physique Th\'eorique UMR 8627,\\
 Universit\'e Paris-Sud 11, Bat. 210, 91405 Orsay Cedex, France}

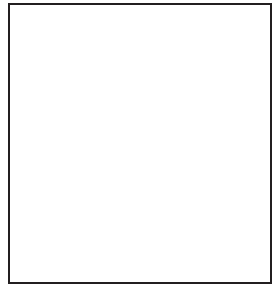
\maketitle\abstracts{
We study the decay rates of the $\mu \rightarrow e \gamma$ and $\tau
\rightarrow \ell \gamma$ transitions in the framework of the type III seesaw model, where fermionic triplets are exchanged to generate neutrino masses. We show that the observation of one of those
decays in planned experiments would contradict bounds
arising from present experimental limits on the $\mu \rightarrow eee$
and $\tau \rightarrow 3 l$ decay rates, and therefore imply that there exist other sources of lepton
flavour violation than those associated to triplet of fermions.}

\section{Introduction}

The Standard Model (SM) has the unique property of conserving flavour in the leptonic sector. However, since the experimental discovery of neutrino oscillations, we know that lepton flavour is violated in the neutrino sector. Neutrinos mass naturally arises within the framework of the seesaw mechanism (via the exchange of heavy fields). In such models flavour violating rare leptonic decays such as $\mu \rightarrow e \gamma$ and $\tau \rightarrow \ell \gamma$ are expected to be relevant. These decays have already been studied  in type I~\cite{muegammatype-I}  and type II~\cite{higgstriplets} seesaw models.  In the following, we study these decays in the framework of the type III seesaw model ~\cite{type-III,ABBGH}, where heavy triplets of fermions are exchanged.

\section{The type-III Seesaw model}

The type-III seesaw model consists in adding $SU(2)_L$ triplets of fermions $\Sigma$, with zero hypercharge, to the SM. At least two triplets are needed to account for the observation of neutrino masses, but in fact only one is sufficient to generate non-vanishing $\ell_1 \rightarrow \ell_2 \gamma$ rate. In the following we will not specify the number of triplets. The heavy fermions are in the adjoint representation of the $SU(2)_L$ group and have a gauge invariant Majorana mass term. The Lagrangian of its interactions reads:
 \begin{equation}
\label{Lfermtriptwobytwo}
{\cal L}=Tr [ \overline{\Sigma} i \slash \hspace{-2.2mm} D  \Sigma ] 
-\frac{1}{2} Tr [\overline{\Sigma}  M_\Sigma \Sigma^c 
                +\overline{\Sigma^c} M_\Sigma^* \Sigma] 
- \tilde{\phi}^\dagger \overline{\Sigma} \sqrt{2}Y_\Sigma L 
-  \overline{L}\sqrt{2} {Y_\Sigma}^\dagger  \Sigma \tilde{\phi}\, ,
\end{equation} 
where $L\equiv (l,\nu)^T$, $\phi\equiv (\phi^+, \phi^0)^T\equiv
(\phi^+, (v+H+i \eta)/\sqrt{2})^T$, $\tilde \phi = i \tau_{2} \phi^*$,
$\Sigma^c \equiv C \overline{\Sigma}^T$ and with, for each fermionic
triplet,
\begin{eqnarray}
\Sigma&=&
\left(
\begin{array}{ cc}
   \Sigma^0/\sqrt{2}  &   \Sigma^+ \\
     \Sigma^- &  -\Sigma^0/\sqrt{2} 
\end{array}
\right), \quad 
\Sigma^c=
\left(
\begin{array}{ cc}
   \Sigma^{0c}/\sqrt{2}  &   \Sigma^{-c} \\
     \Sigma^{+c} &  -\Sigma^{0c}/\sqrt{2} 
\end{array}
\right). 
\end{eqnarray}
After electroweak symmetry breaking, the neutrino mass matrix is given by: $m_\nu=-\frac{v^2}{2} Y_\Sigma^T\frac{1}{M_\Sigma} Y_\Sigma$. The new Yukawa couplings are the source of mixing between the light leptons and the heavy fermions, which, combined with the presence of the Majorana mass term, allow lepton flavour violating processes. Thus, the study of these processes will enable to derive some bounds on the new couplings: $Y_{\Sigma}$ and $M_{\Sigma}$

\section{Flavour changing radiative leptonic decays}
We briefly describe the main steps of the calculation of the $\mu\rightarrow e$ rate. The $\tau$ decay rates will straightforwardly follow. The on-shell transition $\mu \rightarrow e \gamma$ is a magnetic transition, and it can be written, in the limit $m_e\rightarrow 0$, as:
\begin{eqnarray}
T\left(\mu\rightarrow e \gamma\right)=A\times\overline{u_e}\left(p-q\right)\left[iq^{\nu}\varepsilon^{\lambda}\sigma_{\lambda\nu}\left(1+\gamma_5\right)\right]u_{\mu}\left(p\right)\, ,
\end{eqnarray}
where $\varepsilon$ is the polarization of the photon, $p_{\mu}$ the
momentum of the incoming muon, $q_{\mu}$ the momentum of the
outgoing photon and $\sigma_{\mu\nu}=\frac{i}{2}\left[\gamma_{\mu},\gamma_{\nu
}\right]$.
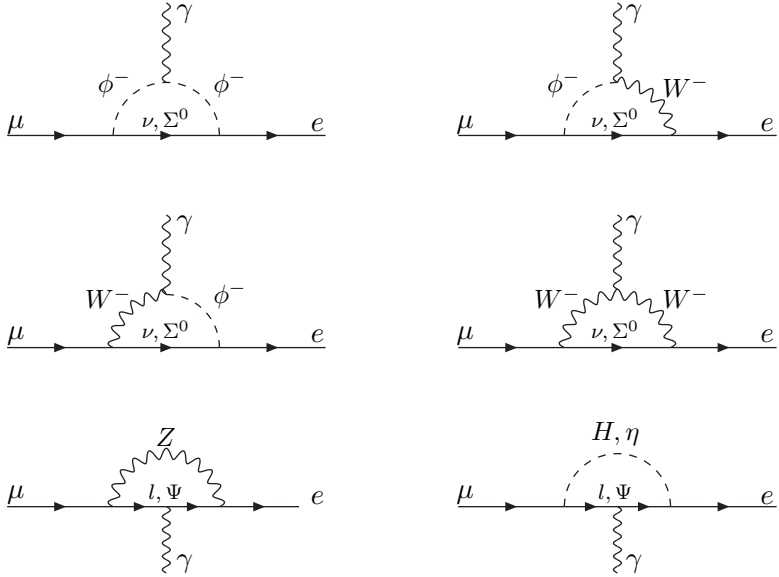
\begin{figure}[t]
\centering
\vspace{15mm}
\begin{picture}(330,150)(0,-100)
\ArrowLine(10,80)(50,80)
\ArrowLine(50,80)(90,80)
\ArrowLine(90,80)(130,80)
\DashCArc(70,80)(20,0,180){3}
\Photon(70,100)(70,130){1.5}{5}
\Text(10,81)[bl]{$\mu$}
\Text(130, 82)[br]{$e$}
\Text(74,130)[tl]{$\gamma$}
\Text(57,95)[br]{\small{$\phi^-$}}
\Text(88,95)[bl]{\small{$\phi^-$}}
\Text(70,82)[b]{\scriptsize{$\nu,\Sigma^0$}}
\ArrowLine(180,80)(220,80)
\ArrowLine(220,80)(260,80)
\ArrowLine(260,80)(300,80)
\DashCArc(240,80)(20,90,180){3}
\PhotonArc(240,80)(20,0,90){2}{5.5}
\Photon(240,100)(240,130){1.5}{5}
\Text(180,81)[bl]{$\mu$}
\Text(300, 82)[br]{$e$}
\Text(244,130)[tl]{$\gamma$}
\Text(227,95)[br]{\small{$\phi^-$}}
\Text(258,95)[bl]{\small{$W^-$}}
\Text(240,82)[b]{\scriptsize{$\nu,\Sigma^0$}}
\ArrowLine(10,0)(50,0)
\ArrowLine(50,0)(90,0)
\ArrowLine(90,0)(130,0)
\DashCArc(70,0)(20,0,90){3}
\PhotonArc(70,0)(20,90,180){2}{5.5}
\Photon(70,20)(70,50){1.5}{5}
\Text(10,1)[bl]{$\mu$}
\Text(130, 2)[br]{$e$}
\Text(74,50)[tl]{$\gamma$}
\Text(57,15)[br]{\small{$W^-$}}
\Text(88,15)[bl]{\small{$\phi^-$}}
\Text(70,2)[b]{\scriptsize{$\nu,\Sigma^0$}}
\ArrowLine(180,0)(220,0)
\ArrowLine(220,0)(260,0)
\ArrowLine(260,0)(300,0)
\PhotonArc(240,0)(20,0,180){2}{10.5}
\Photon(240,22)(240,50){1.5}{5}
\Text(180,1)[bl]{$\mu$}
\Text(300,2)[br]{$e$}
\Text(244,50)[tl]{$\gamma$}
\Text(227,15)[br]{\small{$W^-$}}
\Text(258,15)[bl]{\small{$W^-$}}
\Text(240,2)[b]{\scriptsize{$\nu,\Sigma^0$}}
\ArrowLine(10,-60)(50,-60)
\ArrowLine(50,-60)(70,-60)
\ArrowLine(70,-60)(90,-60)
\ArrowLine(90,-60)(120,-60)
\PhotonArc(70,-60)(20,0,180){2}{10.5}
\Photon(70,-60)(70,-85){1.5}{5}
\Text(10,-59)[bl]{$\mu$}
\Text(130, -59)[br]{$e$}
\Text(74,-85)[bl]{$\gamma$}
\Text(70,-37)[b]{\small{$Z$}}
\Text(70,-59)[b]{\scriptsize{$l,\Psi$}}
\ArrowLine(180,-60)(220,-60)
\ArrowLine(220,-60)(240,-60)
\ArrowLine(240,-60)(260,-60)
\ArrowLine(260,-60)(300,-60)
\DashCArc(240,-60)(20,0,180){3}
\Photon(240,-60)(240,-85){1.5}{5}
\Text(180,-59)[bl]{$\mu$}
\Text(300, -59)[br]{$e$}
\Text(244,-85)[bl]{$\gamma$}
\Text(240,-37)[b]{\small{$H,\eta$}}
\Text(240,-59)[b]{\scriptsize{$l,\Psi$}}
\end{picture}
\caption{Diagrams contributing to the $\mu\rightarrow e \gamma$ transition. $\phi^\pm,\, \eta$
denote the three Goldstone bosons associated with the $W^-$ and $Z$ bosons. $H$ stands for the
physical Higgs boson, and $\Psi=\Sigma^{+c}_{R}+\Sigma^{-}_{R}$.}
\label{muegammatypeIII}
\end{figure}
The fourteen diagrams contributing to these decays are shown in Fig.~1. The details of the calculation can be found in the related paper~\cite{Abada:2008ea}. In the limit
$M_\Sigma\gg M_W$, at ${\cal O}((\frac{Y_\Sigma v}{M_{\Sigma}})^2)$,
the total amplitude is given by:
\begin{eqnarray}
\label{totalamplitude}
T\left(\mu\rightarrow e\gamma\right)&=&
i\frac{{G^{SM}_F}}{\sqrt{2}}\frac{e}{32\pi^2}m_{\mu}\overline{u_e}\left(p-q\right)\left(1+\gamma_5\right)i \sigma_{\lambda\nu} \varepsilon^{\lambda} q^\nu u_{\mu}\left(p\right)\nonumber\\
&\times& \left\{\left(\frac{13}{3}+C\right)\epsilon_{e\mu}-\sum_i x_{\nu_i}\left(U_{PMNS}\right)_{ei}\left(U_{PMNS}^\dagger\right)_{i\mu}\right\}\, ,
\end{eqnarray}
where $C=-6.56$, $\epsilon=\frac{v^2}{2}Y_\Sigma^\dagger M^{-2}_\Sigma Y_\Sigma$ and $x_{\nu_{i}}\equiv \frac{m^2_{\nu_i}}{M_W^2}$. The first part of the amplitude correspond to the contribution of the fermionic triplet, while the second one is the usual contribution from neutrino mixing (suppressed by a GIM cancellation). The branching ratio then reads :
\begin{equation}
\label{brratio}
Br\left(\mu\rightarrow e\gamma\right)=\frac{3}{32}\frac{\alpha}{\pi}\left|\left(\frac{13}{3}+C\right)\epsilon_{e\mu}-\sum_i x_{\nu_i}\left(U_{PMNS}\right)_{ei}\left(U_{PMNS}^\dagger\right)_{i\mu}\right|^2\, .
\end{equation}
$\tau \rightarrow l \gamma$ decays can be obtained from
Eq.~(\ref{brratio}) by replacing $\mu$ by $\tau$, $e$ by $l$ and by
multiplying the obtained result by~\cite{PDG} $Br(\tau \rightarrow e \nu_\tau
\bar{\nu}_e)=(17.84 \pm 0.05) \cdot 10^{-2}$.
Since the neutrino mixing contribution is extremely suppressed, it cna be neglected when compared to the present experimental bounds on the branching ratios~\cite{PDG,Hayasaka:2007vc}. This allow us to convert these experimental bounds into bounds on the $\epsilon_{\alpha\beta}$ coefficients :
\begin{eqnarray}
|\epsilon_{e \mu}|&=&
\frac{v^2}{2}\,|Y_\Sigma^\dagger\frac{1}{M_\Sigma^\dagger}\frac{1}{ M_\Sigma}Y_{\Sigma}|_{\mu e}\ \leq 1.1 \cdot 10^{-4}
\label{bound1C},\\
|\epsilon_{\mu \tau}|&=&
\frac{v^2}{2}\,|Y_\Sigma^\dagger\frac{1}{M_\Sigma^\dagger}\frac{1}{ M_\Sigma}Y_{\Sigma}|_{\tau \mu}\ \leq 1.5 \cdot 10^{-2}
\label{bound2C},\\
|\epsilon_{e \tau }|&=&
\frac{v^2}{2}\,|Y_\Sigma^\dagger\frac{1}{M_\Sigma^\dagger}\frac{1}{ M_\Sigma}Y_{\Sigma}|_{\tau e}\ \leq 2.4 \cdot 10^{-2}\, .
\label{bound3C}
\end{eqnarray}

\section{Comparison to $\mu\rightarrow eee$ and $\tau\rightarrow3\ell'$ decays}

The presence of heavy fermions not only allows for one-loop lepton flavour violating processes, but also for tree level decays such as $\mu\rightarrow eee$ and $\tau\rightarrow3\ell'$. The latter branching ratios have already been calculated in the type III seesaw model~\cite{ABBGH}. It turns out that the bounds obtained from these decays exactly apply on the same parameters $\epsilon$ as those obtained in Eqs.~(\ref{bound1C})-(\ref{bound3C}). To understand this property let us study the example of $\mu\rightarrow e \gamma$ and  $\mu\rightarrow eee$ . In both cases one wants to link a muon to an electron with a same fermionic line. The only way to achieve this is to mix a muon and an electron with a fermionic triplet. This implies that the flavour structure of the $\mu$-to-$e$ fermionic line is the same in both processes. Regarding the couplings, there is only one way to combine two Yukawa couplings and two inverse $M_{\Sigma}$ mass matrices to induce a $\mu$-to-$e$ transition along a fermionic line: $\epsilon_{e\mu}$. This relation between the two types of decays implies that the ratios of these branching ratios are fixed :
\begin{eqnarray}
Br(\mu \rightarrow e \gamma)&=&
1.3 \cdot 10^{-3} \cdot Br(\mu \rightarrow eee)\,, \label{relation1C}\\
Br(\tau \rightarrow \mu \gamma)&=&
1.3 \cdot 10^{-3} \cdot Br(\tau \rightarrow \mu \mu \mu)=
2.1 \cdot 10^{-3}\cdot Br(\tau^-\rightarrow e^- e^+\mu^-)\,,
\label{relation2C}\\
Br(\tau \rightarrow e \gamma)&=&
1.3 \cdot 10^{-3} \cdot Br(\tau \rightarrow e e e)\, \,=
2.1 \cdot 10^{-3}\cdot Br(\tau^-\rightarrow \mu^-\mu^+ e^-)\,.
\label{relation3C}
\end{eqnarray}
Since the processes $\ell\rightarrow 3\ell'$ occur at tree level in this model while the $\ell\rightarrow \ell' \gamma$ ones are one-loop, small values of the ratios are expected\footnote{Note that these ratios are obtained in the limit where $M_\Sigma\gg M_{W,Z,H}$. Not working in this limit, these ratios can vary up to one order of magnitude.}.
The $\mu \rightarrow eee$, $\tau \rightarrow eee$ and $\tau
\rightarrow \mu \mu \mu$ decays lead to~\cite{ABBGH}  $|\epsilon_{e \mu}| < 1.1
\cdot 10^{-6}$, $|\epsilon_{\mu \tau}| < 4.9 \cdot 10^{-4}$,
$|\epsilon_{e \tau}| < 5.1 \cdot 10^{-4}$. Those bounds are better than the one obtained in Eqs.~(\ref{bound1C})-(\ref{bound3C}). This means that in this model the tree-level processes will provide the most competitive bounds on the $\epsilon_{\alpha\beta}$ parameters, even if the the experimental limit on the branchng ratios of the radiative decays improves by two order of magnitude. Using the experimental bounds~\cite{PDG,Belle2} $Br(\mu \rightarrow eee) < 1 \cdot
10^{-12}$, $Br(\tau \rightarrow eee) < 3.6 \cdot 10^{-8}$ and $Br(\tau \rightarrow \mu \mu \mu) < 3.2 \cdot
10^{-8}$, one derives predictions for the bounds on branching ratios of the radiative decays : 
\begin{eqnarray}
Br(\mu
\rightarrow e \gamma) &<& 10^{-15}\\
Br(\tau \rightarrow \mu \gamma)& <& 4 \cdot 10^{-11}\\
Br(\tau \rightarrow e \gamma) &< &5 \cdot10^{-11}
\end{eqnarray}
to be compared with experimental bounds~\cite{PDG,Hayasaka:2007vc} :  Br$(\mu \rightarrow e \gamma) <
1.2 \cdot 10^{-11}$, Br$(\tau \rightarrow \mu \gamma) < 4.5
\cdot 10^{-8}$, Br$(\tau \rightarrow e \gamma)
< 1.1 \cdot 10^{-7}$~\cite{PDG}.

\section{Conclusion}
In our work we were lead to the conclusion that the observation of one leptonic
radiative decay in the upcoming experiments will rule out the
seesaw mechanism with only fermion triplets. Indeed this would contradict bounds
arising from present experimental limits on the $\mu \rightarrow eee$
and $\tau \rightarrow 3 l$ decay rates, and therefore imply that there exist other sources of lepton
flavour violation than those associated to triplet of fermions.

\section*{Acknowledgments}
I would like to thanks the organisers of the XLIII Rencontres de Moriond for giving me the opportunity to give a talk in this conference and for the enjoyable atmosphere. This work was funded in part by a grant from the European Union "Marie Curie" program, and from the Agence Nationale de Recherche ANR through the project JC05-43009-NEUPAC.

\end{document}